\documentclass[conference]{IEEEtran}

\makeatletter

\def\ps@IEEEtitlepagestyle{%
  \def\@oddfoot{\mycopyrightnotice}%
  \def\@evenfoot{}%
}
\def\mycopyrightnotice{%
  {\footnotesize XXX-X-XXXX-XXXX-X/XX/\$XX.00~\copyright~20XX IEEE\hfill}
  \gdef\mycopyrightnotice{}
}

\usepackage{blindtext}
\usepackage{eso-pic}
\IEEEoverridecommandlockouts
\usepackage{cite}
\usepackage{amsmath,amssymb,amsfonts}
\usepackage{algorithm}
\usepackage{algpseudocode}
\usepackage{graphicx}
\usepackage{textcomp}
\usepackage{xcolor}
\usepackage{multirow}
\usepackage{url}
\def\BibTeX{{\rm B\kern-.05em{\sc i\kern-.025em b}\kern-.08em
    T\kern-.1667em\lower.7ex\hbox{E}\kern-.125emX}}
    
\usepackage{eso-pic}

\newcommand\correspondingauthor{\thanks{Corresponding author: Iztok Fister Jr. (e-mail: iztok.fister1@um.si).}}
    
\begin{document}
\title{\vspace*{1cm} Profiling the carbon footprint of performance bugs
}

\author{\IEEEauthorblockN{1\textsuperscript{st} Iztok Fister Jr.\correspondingauthor}
\IEEEauthorblockA{\textit{University of Maribor}\\
Maribor, Slovenia \\
iztok.fister1@um.si}
\and
\IEEEauthorblockN{2\textsuperscript{nd} Du\v{s}an Fister}
\IEEEauthorblockA{\textit{University of Maribor}\\
Maribor, Slovenia \\
dusan.fister@um.si}
\and
\IEEEauthorblockN{3\textsuperscript{th} Vili Podgorelec}
\IEEEauthorblockA{\textit{University of Maribor}\\
Maribor, Slovenia \\
vili.podgorelec@um.si}
\and
\IEEEauthorblockN{4\textsuperscript{th} Iztok Fister}
\IEEEauthorblockA{\textit{University of Maribor}\\
Maribor, Slovenia \\
iztok.fister@um.si}
}

\maketitle
\begin{abstract}
Much debate nowadays is devoted to the impacts of modern information and communication technology on global carbon emissions. Green information and communication technology is a paradigm creating a sustainable and environmentally friendly computing field that tries to minimize the adverse effects on the environment. Green information and communication technology are under constant development nowadays. Thus, in this paper, we undertake the problem of performance bugs that, until recently, have never been studied so profoundly. We assume that inappropriate software implementations can have a crucial influence on global carbon emissions. Here, we classify those performance bugs and develop inappropriate implementations of four programs written in C++. To mitigate these simulated performance bugs, measuring software and hardware methods that can estimate the increased carbon footprint properly were proposed.
\end{abstract}



\section{Introduction}
We find ourselves in an era marked by turbulence, where the ominous specters of global warming and excessive carbon emissions loom large in our collective consciousness~\cite{liu2022monitoring}. Many scientists have long sounded the alarm about the urgent need to curb these emissions to avert catastrophic consequences. However, despite the multitude of dire projections, political movements worldwide often fall short in their efforts to mitigate the impact of global warming effectively~\cite{nascimento2023expanding}.

Modern humans and novel economic systems play a pivotal role in exacerbating the carbon footprint. On one hand, we engage in activities such as deforestation, construction on valuable land, the proliferation of supermarkets, while, on the other, in the extensive use of automobiles and airplanes~\cite{balboni2023economics,sagar1995automobiles}. Moreover, it has become increasingly evident that Information and Communication Technology (ICT), which we rely upon heavily, contributes significantly in increasing the carbon emissions. Considering the vast number of data centers scattered across the globe~\cite{koomey2011growth}, as well as the multitude of machine learning models running incessantly~\cite{dhar2020carbon}, it becomes clear just how profound our environmental impact can be. In addition to all these examples, the personal computers are using electricity, and potentially contribute in increasing the carbon emissions~\cite{somavat2011energy}.

The task of Green ICT is to mitigate the carbon emissions of ICT production, applications, and services~\cite{Taina2010how}. It holds that the ICT systems nowadays produce even 2~\% of global emissions~\cite{kern2015impacts}. The reduction of carbon emissions comes out either directly from the hardware, or directly and indirectly from the software. Indeed, the carbon emission is measured as a Carbon Footprint (CF) that is proportional to the amount of trees needed to absorb the emitted carbon dioxide in a year~\cite{Taina2010how}. 

Several publications have tackled the challenge of enhancing the sustainability of ICT and its associated processes. In a paper authored by Taina~\cite{Taina2010how}, a comprehensive approach was presented for analyzing the carbon footprint associated with software. This study examined each phase of a typical software lifecycle meticulously, quantifying the carbon emissions generated at each step. Additionally, the author shared insightful strategies aimed at minimizing this carbon footprint. Conversely, in paper~\cite{kern2015impacts}, the authors introduced a systematic methodology for quantifying the carbon footprint of a software product throughout its entire lifecycle. Furthermore, they proposed a method for incorporating certain facets of carbon footprint assessment seamlessly into the software development process. The paper also delves into the implications and tools associated with this innovative calculation approach. Thus, this work underscores the significance of energy metrics, and the consideration of carbon footprint implications within the realm of Green Software Engineering.

Performance bugs are unnecessarily inefficient code chunks in software that can cause prolonged execution times and degraded resource utilization~\cite{Chen2022slowcoach}. For instance, an execution time of a program calling a function each time it needs in place of referencing the variable storing the result of the calling function can increase its execution time substantially. The execution time is increased proportionally to the number of function calls. 

The impact of the unnecessary inefficient code chunks has rarely been taken into consideration, especially in the sense of Green ICT. The purpose of the study is to observe the well known performance bug in C++ referring to a vector class, and to show how its inefficient usage can increase the energy consumption (indirectly also carbon emission) of the corresponding algorithm. Vector in C++ reallocates memory when the new elements are added, and no memory is available to hold them. The reallocation would cause significant performance degradation if this occurs too often~\cite{Chen2022slowcoach}.

To simulate performance bug, four different versions of an algorithm were developed that manipulate the big number of elements, either in the vector class or the double linked list. All the algorithms were executed on three different platforms (i.e., a laptop, a Raspberry Pi 3 microcomputer, and an iPad table computer), and compared with each other according to the increased energy consumption measured, depending on the platform either using the Linux system software tool or a power meter capable of measuring the energy consumption on the hardware level.

The motivation of the study was three-fold:
\begin{itemize}
    \item Identifying the performance bugs in software or inappropriate implementation that can have a great influence toward the carbon emission. 
    \item Investigating the influence of the iteration performance bugs on the increased carbon emission by results of four simulation programs developed in C++ programming language. 
    \item Searching for methods of how to measure the influence of performance bugs.
\end{itemize}
Although it can be found three performance bugs in the literature~\cite{Chen2022slowcoach} (i.e., iteration, enumeration, and deadlock performance bugs), here, we are focused on the first kind only. As a result, the main contributions of the study are as follows:
\begin{itemize}
    \item Simulating the performance bugs by four different algorithms written in C++.
    \item Measuring an increasing carbon footprint caused by the simulation.
    \item Showing that the proposed measuring methods can be applied to estimate the increased carbon footprint due to the simulated performance bugs.
\end{itemize}
In general, the main novelty of the proposed method is to link the identification of performance bugs with measuring the increased carbon footprint they cause. Thus, the domain of software engineering is integrated with the domain of green computing.

The structure of the paper is as follows: Section~\ref{sec:2} discusses potential ways in which to measure the carbon footprint on various digital computers. In Section~\ref{sec:3}, the concept of performance bugs is explained briefly. The performed experiments and the obtained results are the subjects of Section~\ref{sec:4}. Section~\ref{sec:5} concludes the paper and outlines the plausible directions for the future work.

\section{Measuring the carbon footprint of computers}~\label{sec:2} 
A carbon footprint is the amount of greenhouse gases (e.g., carbon dioxide and methane) that are generated by our actions~\cite{nature2023calculate}. The carbon footprint is measured in units of CO$_2$ per unit (CO$_2$e). Actually, most things in the world contribute to carbon emissions, i.e., has its carbon footprint~\cite{berners2022carbon}. Electricity belongs today to the primary source of energy. Moreover, energy consumption is increasing crucially. As the production of electricity increases, the carbon emissions rise simultaneously. However, electricity can be produced based on different sources (e.g., coal, oil, gas, nuclear and renewable) that contribute a different level of carbon emissions. For instance, the low-carbon sources, like nuclear and renewable, are more environmentally friendly than the coal that belongs to the high-carbon sources. 

Typically, each country produces electricity from sources of different levels of carbon emissions. Therefore, the carbon footprint of electricity varies from country to country. The carbon footprint for electricity in the United States is estimated to be 0.65~kg~CO$_2$e. 

The energy consumption needs to be estimated in order to determine the carbon footprint of computers. Although the electricity is not the only source of carbon emissions by computers~\cite{berners2022carbon}, it is a good approximation for calculation of the carbon footprint. Indeed, there are two ways to determine the energy consumption of computers, i.e., software and hardware tools. However, the known software tools only measure power consumption on laptops when running on a battery. On the other hand, for desktop or server machines the only current solution is an electronic power-meter that plugs into the mains socket. In our study, the \textit{powertop} utility was examined among the software tools, and the AVHzY ct-3 power-meter among the hardware tools. The characteristics of both tools are discussed in the remainder of the paper.

\subsection{Powertop utility on Linux}
The purpose of the \textit{powertop} utility is to analyze and manage power consumption on laptops using battery power. The tool is able to display and export reports about the estimated discharge rate, and statics about processors, devices, kernel, timer, and interrupt handler behavior. It also lets us tune some kernel parameters easily on the fly, in order to maximize the battery life. 

Before operating the tools needs to be calibrated, where, during the process, the power engine adjusts to the specific computer environment in order to take the accurate power measures. In operation mode, the laptop must be on battery power only. The \textit{powertop} takes measurements at 20~seconds intervals by default. Obviously, the interval can be set arbitrarily by the user. 

Indeed, the process of discharging the battery is strongly non-linear. Consequently, measurements taken at full battery capacity under a similar strain can differ from those taken when the capacity of the battery is not full. As a result, we need to ensure that the battery is at full capacity at the beginning of each experiment. 

\subsection{Power meter AVHzY CT-3}
A power-meter measures electrical power in Watts. For measuring computer power, electronic Watt-meters are used that are capable of small power measurements, or of power measurements at frequencies beyond the range of dynamometer-type instruments. 

\begin{figure}[htb]
    \centering
    \includegraphics[scale=0.64]{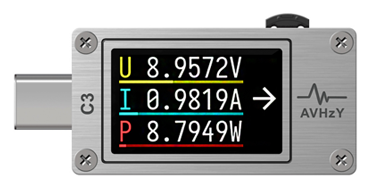}
    \caption{Power meter AVHzY ct-3.}
    \label{fig:avhzy}
\end{figure}

The AVHzY CT-3 power-meter (Fig.~\ref{fig:avhzy}) measures active electrical power while connected directly to the electricity network, and works in $0-26$~V of voltage and $0-6$~A of 
current ranges. The device is connected to the electricity network via a 230~V plug adapter with a USB port. It measures the consumer that is connected to it via an output USB port. Furthermore, the power-meter also enables connecting the external power via a USB-C input port. 

The main advantage of the power-meter is represented by the USB-C output port, to which the personal computer can be connected, on which the powerful PC software can be installed for monitoring power consumption simultaneously. The software is dedicated for data logging up to 1000sps, viewing the VBUS ripple, and diagnosing the devices. Using this PC software, the power consumption can be monitored online at a high level of accuracy.

\section{Performance bugs}~\label{sec:3}
Performance bugs typically increase the time complexity of the algorithm due to coding the inefficient code chunks. Obviously, this inefficient coding is programmed by the programmer unintentionally, but has a crucial impact on the performance behavior of the algorithm.  For instance, allocations of heap memory are performed with a \textit{malloc} function in standard C and with \textit{new} function in C++. If low-level system allocations are needed, the custom allocators, e.g., kalloc in Linux kernel, or ALLOC and xmalloc in gnulib, are available for vector reallocation. Consequently, performance improvements are clearly observable by calling the low-level allocation function.

In order to show, how the performance bugs affect the performance of the algorithm, different applications of the vector class in C++ were taken into consideration. Indeed, the memory is allocated on demand by the class, and is deallocated at the very least when the vector's destructor is called. When no memory is available to hold the new allocated elements in a heap, the vector in C++ needs to reallocate memory. These reallocations can be simulated as a performance bug by mass insertion of the new head vector's elements by its member function \textit{push\_back}, and mass deletion of the last vector's element by the member function \textit{pop\_back}.

In line with this, the following four different algorithms were developed:
\begin{itemize}
    \item Vector (i.e., the simulation of the performance bug),
    \item Raw (i.e., avoiding the performance bug version~I),
    \item Array (i.e., avoiding the performance bug version~II),
    \item double linked List (i.e., avoiding the performance bug version~III).
\end{itemize}
obtaining the same results in different ways, of course. The task of each algorithm is simply to initialize the array of long integer elements with their sequence numbers in the interval $[0,99]$, and, then, iterate deleting the last element of the array and adding the next element in the sequence of numbers to the head of the array. Finally, the elements of the array are ordered in descend order starting with the the first number that is equal to the maximum number of iterations, until the number that is for 100 elements lower than the maximum. Thus, two representations of the array are applied, i.e., vector class and array data structure in C++. Thus, it is expected that algorithms using the elementary data structures (e.g., Array and Link) would be more efficient than those using the more abstract Vector class (e.g., Vector and Raw).

The pseudo-code of the algorithm Vector is illustrated in Algorithm~\ref{alg:1}, 
\begin{algorithm}[htb]
\caption{Algorithm Vector}
\label{alg:1}
\begin{algorithmic}
\State vector $<\text{long int}>$ vec \Comment{vector class C++}
\For {i=0L \textbf{to} 100L}
\State vec.push\_back(i)\Comment{Initialization of vector}
\EndFor
\For {i=0L \textbf{to} MAX\_ITER}\Comment{Program loop}
\State vec.pop\_back()\Comment{Delete the last element}
\State vec.insert(head, i+1)\Comment{Insert new as the first element}
\EndFor
\end{algorithmic}
\end{algorithm}
from which it can be seen that this uses the vector class functions \textit{push\_back}, \textit{pop\_back}, and \textit{insert} for adding the last, removing the last and inserting the first element into/from the vector variable \textit{vec}.

Algorithm Raw is implemented according to the pseudo-code depicted in Algorithm~\ref{alg:2}.
\begin{algorithm}[htb]
\caption{Algorithm Raw}
\label{alg:2}
\begin{algorithmic}
\State vector $<\text{long int}>$ vec \Comment{vector class C++}
\For {i=0L \textbf{to} 100L}
\State vec.push\_back(i)\Comment{Initialization of vector}
\EndFor
\For {i=0L \textbf{to} MAX\_ITER}\Comment{Program loop}
\For {j=99 \textbf{to} 1 \textbf{step} -1}\Comment{Exchange elements}
\State vec[j] = vec[j-1]
\EndFor
\State vec[0] = i+100L
\EndFor
\end{algorithmic}
\end{algorithm}
As evident from the pseudo-code, the array of integer elements is defined as a class of long integer, where the initialization is performed in the same way as by the algorithm Vector (i.e., using the \textit{push\_back} function call). However, manipulation of the vector elements is developed in a more elementary way: At first, the whole array are reassigned sequentially element by element backward, while the first element is adopted with the number proportional to the current iteration (i.e., variable $i$).

Algorithm Array, presented in pseudo-code Algorithm~\ref{alg:3}, applies the array C++ data structure for representation of elements of type long integer.
\begin{algorithm}[htb]
\caption{Algorithm Array}
\label{alg:3}
\begin{algorithmic}
\State long int vec[100]\Comment{array C++ data structure}
\For {i=0L \textbf{to} 100L}
\State vec[i] = i \Comment{Initialization of vector}
\EndFor
\For {i=0L \textbf{to} MAX\_ITER}\Comment{Program loop}
\For {j=99 \textbf{to} 1 \textbf{step} -1}
\State vec[j] = vec[j-1]
\EndFor
\State vec[0] = i+100L
\EndFor
\end{algorithmic}
\end{algorithm}
As is evident from the pseudo-code, the implementation is similar to the implementation of the Raw algorithm, except in the declaration of the array variable \textbf{vec}. Here, the array data structure in C++ is employed in place of using vector class. The motivation behind implementing the algorithm is to show, which potential overhead brings the introduction of the more abstract class vector over the more elementary data structure array in C++.

Finally, the algorithm List implements the double linked list data structure in C++. The advantage of this algorithm is that the functions of the high-level vector class are replaced with the low-level functions implemented by its own, which are able to optimize the algorithm's behavior in the sense of speed and space. The double linked list uses the data structure as illustrated in Algorithm~\ref{alg:4},
\begin{algorithm}[htb]
\caption{Double linked List data structure}
\label{alg:4}
\begin{algorithmic}
\State struct str\_list \{
\State \qquad long int num;
\State \qquad str\_list* prev;
\State \qquad str\_list* next;
\State \} *list;\Comment{double linked list}
\end{algorithmic}
\end{algorithm}
from which it can be seen that this consists of two pointers \textbf{prev} and \textbf{next}, and the long integer variable \textbf{num}. 

The pseudo-code of the algorithm List is presented in Algorithm~\ref{alg:5}, 
\begin{algorithm}[htb]
\caption{Algorithm double linked List}
\label{alg:5}
\begin{algorithmic}
\State struct str\_list* first = NULL
\State struct str\_list* last = NULL
\For {i=99L \textbf{to} 1L \textbf{step} -1L}
\State first = add\_new(first, last, i) \Comment{Initialization of vector}
\If {i == 99}
\State last = first
\EndIf
\EndFor
\For {i=0L \textbf{to} MAX\_ITER}\Comment{Program loop}
\State last$->$num = i + 100L
\State first = last
\State last = last$->$prev
\EndFor
\end{algorithmic}
\end{algorithm}
from which it can be considered that the elements are entered into the double linked list by calling the function \textit{add\_new}. The implementation of the function is straightforward, and demands only to allocate the new item in the heap, initialize it, and ensure the proper forward and backward linking. Therefore, the pseudo-code of the function is not presented in the paper. Interestingly, the effect of element temptation is achieved by putting the current iteration number into the first element and tying the other 99 elements for one. This whole task can be performed simply by modifying the value of the pointers \textbf{first} and \textbf{last} (i.e., no reallocation is needed).

\section{Experiments and results}~\label{sec:4}
The purpose of our experimental work was to show what amount of carbon footprint can be expected due to performance bugs. In line with this, the four implemented algorithms (i.e., Vector, Raw, Array and List)~\footnote{\url{https://codeberg.org/firefly-cpp/green-ict-benchmarks}} were compared according to their carbon footprint, measured on three different computer platforms, i.e.,:
\begin{itemize}
    \item with software tools on a laptop,
    \item with a power-meter on a Raspberry Pi 3,
    \item with a power-meter on an Apple iPad.
\end{itemize}
In the remainder of the paper, the results of all three measurements are discussed in detail. 

\subsection{Measuring the carbon footprint on a laptop}
The effective electrical energy consumption was measured on a laptop using battery power with \textit{powertop} software utility under the Linux operating system. The four implemented test algorithms were running on the mentioned computer platforms with the maximum number of iterations set to $\text{MAX\_ITER}=10^{10}$. All the programs were performed independently over 10 runs and the average measures were taken into consideration.

The characteristics of the laptop battery are illustrated in Table~\ref{tab:1}.
\begin{table}[htb]
    \caption{Laptop battery characteristics.}
    \label{tab:1}
    \centering
    \begin{tabular}{l|r}
    Specification & Description \\\hline
    Type & HP ProBook 470 G3 \\
    Amp-hour capacity & 3 Ah \\
    Watt-hour capacity & 44 Wh \\
    Voltage & 14.8 V \\
    Cell type & 4 cell Lithium-Ion \\\hline
    \end{tabular}
\end{table}
Let us mention that the laptop battery was at full capacity before each start of the particular experiment.

The results of the experiments are depicted in Table~\ref{tab:2} that is divided into the following columns: the "Init DR" represents the average initial Discharging Rate (DR) before and after running the algorithm, the "Running DR" is the DR under the strain, the "Alg. DR" refers to the DR caused by the running algorithm, the "Time" measures the average execution time of the algorithm, the "Energy" denotes the energy used as a product of algorithm DR by the time, and the "Carbon footprint" estimates the carbon footprint as a product of CO$_2$ per unit emitted by one kWh (i.e., 0.65 CO$_2$e/kWh). 
\begin{table*}[htb]
    \caption{Carbon footprint obtained on the laptop by different C++ algorithms.}
    \label{tab:2}
    \centering
    \begin{tabular}{l|rrrrrr}
    \multirow{2}{*}{Algorithm} & Init DR & Running DR & Alg. DR & Time & Energy & Carbon footprint \\
        & [W] & [W] & [W] & [sec] & [Wh] & [CO$_2$e/Wh] \\\hline
        Vector	&	9.288	&	15.953	&	6.665	&	198.496	&	22.051	&	14.333	\\
        Raw	    &	9.010	&	16.142	&	7.132	&	195.606	&	23.250	&	\textbf{15.113}	\\
        Array	&	9.532	&	16.221	&	6.689	&	177.381	&	19.776	&	12.854	\\
        List	&	7.850	&	15.138	&	7.288	&	16.081	&	1.953	&	1.270	\\\hline
    \end{tabular}
\end{table*}
As indicated from the table, the Raw algorithm produced the higher carbon footprint among the four algorithms. As expected, the List algorithm was the most green in the sense of Green ITC.

\subsection{Measuring the carbon footprint with a power-meter on a Raspberry Pi 3}
Raspberry Pi is a series of small single-board computers (SBCs) based on an ARM processor that represents an all in one computer. Due to its affordability, it is suitable for using in teaching basic computer science in schools. In our study, the Raspberry Pi was used that was equipped with a 32-bit ARM processor, 1~GB memory, 256~GB SSD, DVI video port, Ethernet and Wireless LAN. The Raspbian OS was installed on the device. The main advantage of this all in one computer represents its power supply over a USB-C port. This means that it is able to be monitored using the already mentioned AVHzY CT-3 power-meter. Let us mention that the maximum number of iterations was set to $\text{MAX\_ITER}=10^8$.

\begin{figure*}
    \centering
    \includegraphics[scale=0.34]{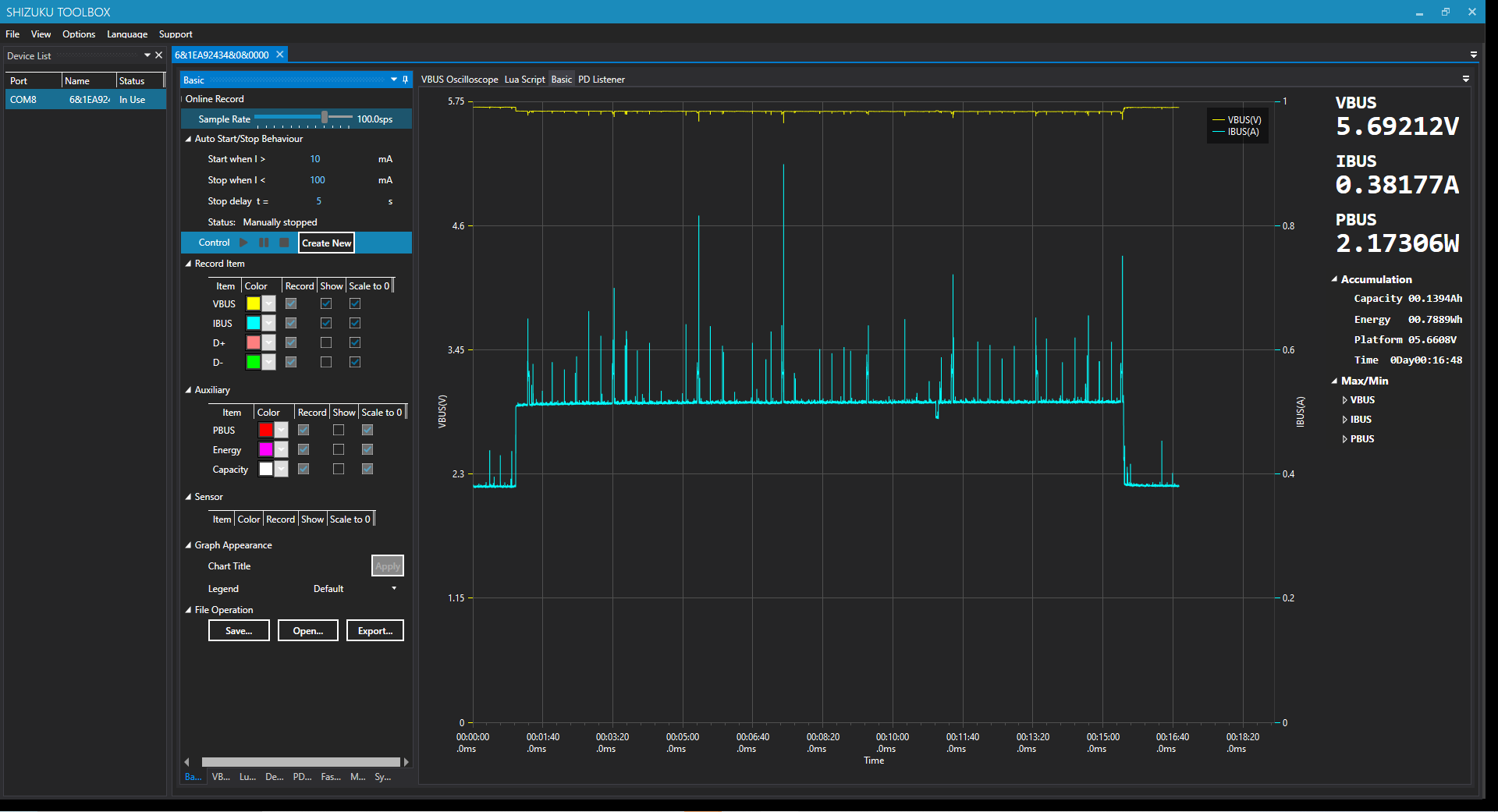}
    \caption{Measuring protocol on AVHzY CT-3 by Array algorithm strain.}
    \label{fig:raspberry}
\end{figure*}

An example of the measuring protocol obtained by measuring the power consumption by executing the Array algorithm by the PC is illustrated in Fig.~\ref{fig:raspberry}.
The curve in the figure illustrates the VBUS ripple that reflects a reaction of the Raspberry Pi in the strain caused by executing the Array algorithm. As can be observed, the strain is demonstrated as a step function denoting the increasing of the voltage. Interestingly, the sudden withdrawal of voltage is a typical consequence of some interrupts caused by the algorithm itself (e.g., by the output of control messages), different I/O actions (e.g., moving the mouse) or system actions. Let us mention that no other programs were active during the experimental work. 

The results of measuring the carbon footprint are depicted in Table~\ref{tab:3}. Actually, all the experiments consisted of three phases: (1) measuring the inactivity before the strain, (2) measuring the strain activity, and (3) measuring the inactivity after the strain. Typically, the duration of both inactivity phases was approximately one minute, while the strain activity phase denotes the effective power obtained during executing the particular algorithm. In the table, the row "Average inactivity" denotes the average carbon footprint in CO$_2$/Wh measured during both inactivity phases, the row "Total strain" is the total carbon footprint measured during the strain phase, while the row "Algorithm strain" refers to the foot print caused by executing the algorithm (i.e., simply the difference between the total and the average strain).
\begin{table}
    \caption{Results of measuring the carbon footprint.}
    \label{tab:3}
    \centering
    \begin{tabular}{l|rrrr}
        Carbon footprint & Vector & Raw & Array & List \\\hline
        Average inactivity 	&	3.299	&	3.242	&	3.273	&	3.637	\\
        Total strain 	&	12.860	&	63.659	&	27.496	&	8.211	\\
        Algorithm strain 	&	9.561	&	60.417	&	24.223	&	\textbf{4.574}	\\\hline
    \end{tabular}
\end{table}
As evident from the table, the double linked list implementation is the most environmentally sustainable in the sense of emitting the lowest carbon footprint. The Vector implementation of the algorithm was better than the rest of the algorithms in the tests.

\subsection{Measuring the carbon footprint with a power meter on an Apple iPad}
The Apple iPad prefers the foreground applications in order to ensure users the online response. Therefore, terminations are part of the application life-cycle, when the system terminates the long-term running process. In the case of running the C++ application, this termination is followed with issuing the message "Too much resources..." by the iOS after approximately 20~sec. Although the iPad allows running a long-term application in the background asynchronously, our goal was to indicate the carbon footprint of the foreground processes running on the iPad. In line with this, the maximum number of iterations need to be reduced significantly to $\text{MAX\_ITER}=10^7$. The specifications of the iPad were as follow: model number MYLA2HC/A, iPad OS version 16.7, processor $2\times\text{Vortex}$, 3~GB operating memory, and 32~GB built-in memory. 

\begin{figure*}[htb]
    \centering
    \includegraphics[scale=0.91]{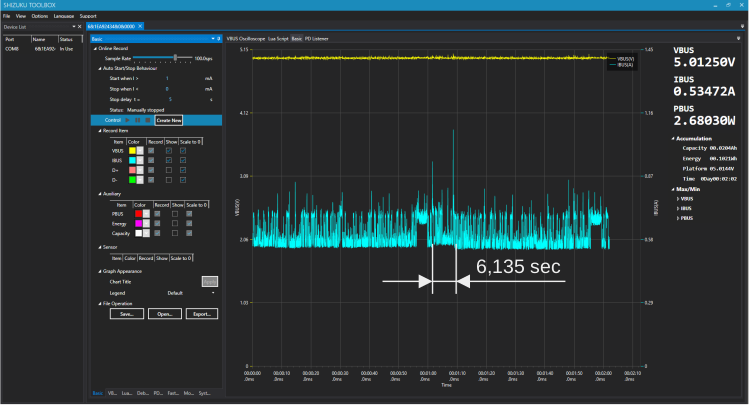}
    \caption{Measuring protocol on AVHzY CT-3 by Array algorithm strain on iPad.}
    \label{fig:ipad}
\end{figure*}

The measuring protocol obtained by the Array algorithm strain on the iPad using the power-meter AVHzY CT-3 was presented in Fig.~\ref{fig:ipad}, 
from which it can be concluded that the supplement of the Array algorithm to the total power consumption (i.e., the VBUS ripple) was not easy to observe. Therefore, this supplement of 6.135~seconds is denoted in the figure by arrows. When one compares the curve with those presented in Fig.~\ref{fig:raspberry} obtained on the Raspberry Pi, he/she can conclude that the iPad system is very agile when handling a lot of active processes simultaneously. Although frequent withdrawal of voltage can be indicated in the figure, the running application left its carbon footprint clearly.

The measured carbon footprint is depicted in Table~\ref{tab:4}, 
\begin{table}[htb]
    \caption{Apple iPad}
    \label{tab:4}
    \centering
    \begin{tabular}{l|rrrr}
        Carbon footprint    &	Vector  &	 Raw 	&	 Array 	&	 List 	\\\hline
        Average inactivity 	&	6.087	&	6.121	&	5.808	&	5.975	 \\
        Total strain 	    &	0.855	&	1.289	&	0.637	&	0.824	 \\
        Algorithm strain 	&	0.123	&	0.516	&	0.317	&	\textbf{0.075}	\\\hline 
    \end{tabular}
\end{table}
from which it is evident that the carbon footprint by the algorithms in test was lower due to the lower execution time, but the relations between the algorithms remained the same as in the last experiment, i.e., the double linked list left the lesser carbon footprint, while the Raw algorithm was distinguished as the worst environment pollutant.

\subsection{Discussion}
As evident from the performed experimental work, measuring the power consumption is more accurate using the hardware device (i.e., the power-meter) than with software tools (i.e., the \textit{powertop}). Although the measurements using \textit{powertop} were taken at 1~second intervals, less than one-third could be captured in one minute (i.e., 16/60). The reason behind the behavior needs to be searched for in the slow communication of the tool with the power engine. On the other hand, the communication of the control PC with the power-meter was very fast and accurate. The hardware device is able to transmit up to 100 measurements in one second. This means that even small changes of power consumption can be sensed by the power-meter. However, the basic problem which arose by using the AVHzY CT-3 device with the laptop was that the power-meter device was able to support only devices connected to the power source with the USB port. This means that for measuring power consumption on laptops the more professional equipment is necessary.

In first sight, it seems that measuring the carbon footprint was measured incorrectly, because the measured algorithm was executed on an operating system that handles a lot of the other programs simultaneously and also affected the increased power consumption. However, the number of these processes (e.g., internet explorer, e-mail client, etc.) was minimized during the measuring, while the average of both the so-called inactivity phases (i.e., before and after executing the algorithm) were measured explicitly. Obviously, the best solution is to put the computer in single-user mode, but this option is unfortunately not available to all computers (e.g., an iPad).

It turns out, that the vector in C++ is sensitive on performance problem only, when the data class is used a lot of the time. Interestingly, sequential usage of the array data structure by the algorithm Array was even more efficient than using the built-in functions by the algorithm Vector, when the enormous actions were applied on the class. As expected, the algorithm List, using the classical implementation of double linked list, outperformed all the other algorithms.

\section{Conclusion}~\label{sec:5}
The paper tries to achieve three goals: (1) to estimate how the performance bugs in software influence increasing the carbon footprint, (2) to focus primarily on the iteration performance bugs, and (3) to find methods for measuring the carbon footprint caused by the performance bugs properly. In line with this, four algorithms simulating the iterative performance bugs were considered and run on three platforms (i.e., a laptop, a micro-computer Raspberry Pi, and an Apple iPad). Two methods for measuring the carbon footprint were examined, i.e., using the software tool \textit{powertop} on Linux and the power meter AVHzY CT-3. The experiments revealed that the hardware measurement using the power meter was more accurate than those using the software tools. However, both can adequately identify the increased carbon footprint caused by the simulation.

In summary, the study integrates two domains, i.e., software engineering and green computing. The former can identify the software bugs, while the latter deals with the effects of computing on global warming. In line with this, this study warns of harmful effects caused by performance bugs in the sense of the increased carbon footprint.

As potential directions for the future, searching for cheaper and more accurate solutions should be made for measuring power consumption on computers connected to the 230~V electricity network(e.g., AVHzY AC WiFi Watt Meter). Also, widening the study to include all performance bugs would be welcome.

\bibliography{bibtex.bib}
\bibliographystyle{plain}

\end{document}